# Signal Partitioning Algorithm for Highly Efficient Gaussian Mixture Modeling in Mass Spectrometry

Andrzej Polanski [1], Michal Marczyk[2], Monika Pietrowska[3], Piotr Widlak[3], Joanna Polanska[2*]

[1]Institute of Informatics, Silesian University of Technology, Gliwice, Poland

[2]Institute of Automatic Control, Silesian University of Technology, Gliwice, Poland

[3]Maria Sklodowska-Curie Memorial Cancer Center and Institute of Oncology, Gliwice, Poland

*Corresponding author

E-mail: joanna.polanska@polsl.pl (JP)

# Abstract

Mixture - modeling of mass spectra is an approach with many potential applications including peak detection and quantification, smoothing, de-noising, feature extraction and spectral signal compression. However, existing algorithms do not allow for automatic analyses of whole spectra. Therefore, despite highlighting potential advantages of mixture modeling of mass spectra of peptide/protein mixtures and some preliminary results presented in several papers, the mixture modeling approach was so far not developed to the stage enabling systematic comparisons with existing software packages for proteomic mass spectra analyses. In this paper we present an efficient algorithm for Gaussian mixture modeling of proteomic mass spectra of different types (e.g., MALDI-ToF profiling, MALDI-IMS). The main idea is automatic partitioning of protein mass spectral signal into fragments. The obtained fragments are separately decomposed into Gaussian mixture models. The parameters of the mixture models of fragments are then aggregated to form the mixture model of the whole spectrum. We compare the elaborated algorithm to existing algorithms for peak detection and we demonstrate improvements of peak detection efficiency obtained by using Gaussian mixture modeling. We also show applications of the elaborated algorithm to real proteomic datasets of low and high resolution.

# Introduction

Current computational methodology for processing signals from spectra registered by mass spectrometry (MS) in mixtures of proteins and/or peptides usually involves sequences of signal processing operations organized in a manner leading to the detection and quantification of spectral peaks. When proteomic mass profiles are analyzed and interpreted, spectral peaks are used as features of MS spectra. It is assumed that each spectral peak corresponds to a certain peptide/protein species, and the composition of the mass spectrum carries direct information on composition of the analyzed samples. Currently, there are already more than a dozen algorithms-, either publicly available or as commercial software packages, that enable proteomic MS spectral peak detection and quantification [1-17]. Different algorithms apply different procedures, a different order and/or variants of signal processing operations. Algorithms can also differ with respect to types of proteomic data (e.g. MALDI/SELDI-ToF profiling, MALDI-IMS, LC-MS/MS).

A potentially useful approach to computational processing of proteomic MS spectra is modeling spectral signals by mixtures of component functions. Some results in this area were published in [18-24]. A natural choice for the component functions are Gaussian distribution functions. However, the use of other component functions has also been studied [18]. Several advantages of using mixture modeling for protein MS spectra are highlighted in the referenced studies [18-24]. Using mixture models potentially allows for more accurate peak detection and quantification. In particular, in the cases where there are overlaps between components (peaks), mixture models enable detecting components "hidden" behind others. Components of mixture models of MS spectra are characterized by both positions and shapes (widths), while in most peak detection methods the information on shapes is missing. Fitting a mixture of components model to actual MS spectra allows for achieving higher sensitivity in detecting peaks of low intensity. The method of decomposition of the spectral signal into components can be more robust against disturbances.

Applications of mixture modeling to proteomic MS spectra were researched in [18-24] by analyzing proteomic actual mass spectra, or their fragments, and by conducting experiments involving fitting mixture models to data. A computational model and some exemplary results were presented in [22]. In [19] Kempka and coauthors studied a model based on the biophysical mechanisms of forming peaks in the MALDI ToF MS spectrum, with two Gaussian components corresponding to two sets of ions formed during the peptide ionization stage. Dijkstra and coauthors [18] have proposed an algorithm for fitting a mixture of a uniform distribution, exponential distribution and a number of log normal distributions to SELDI ToF spectra. Wang and coauthors [24] fitted a mixture of polynomial and Gaussian components to fragments of SELDI ToF spectra and used the MCMC (Markov chain Monte Carlo) approach for iterative estimation of mixture parameters. Noy and Fasulo [20] proposed a method of decomposing protein mass spectra with a set of component distributions derived from peptide models expected to be present in the samples. Positions and shapes of Gaussian components were fixed, the model was fitted to the data by iterations involving only component weights (Watson – Nadaraya iterations). Pelikan and Hauskrecht [21] also used predefined components following from characteristics of peptides/proteins expected in the samples. They have fitted model to data by using Bayesian probabilistic model and dynamic programming algorithm. In a recent paper [23] the authors have fitted homoscedastic Gaussian mixture models to small fragments of high resolution spectra and demonstrated its efficiency for MS signal quantification.

While highlighting a potential of application of mixture modeling to proteomic mass spectra, the studies mentioned above did not lead to algorithms capable to perform analyses based on automatic mixture decompositions at the whole spectrum scale. Methods presented in [20] and [21] enable whole spectrum analyses, but require collecting information on peptides expected in the samples, which results in the restriction of its application to analyses of samples with a known peptide composition. The methods presented in [18], [19] and [23] could be applied to analyses at the whole spectrum scale only with a large amount of human processing involving e.g., the appropriate partitioning of spectral lines. Consequently, in the referenced studies the algorithms for modeling proteomic MS signals by mixtures of component functions were not compared to existing algorithms and software packages for peak detection in the sense of performing sufficiently large computational experiments and comparing the values of the detection efficiency indexes.

The standard approach to modeling signals by mixtures of Gaussian component functions is by using expectation – maximization (EM) algorithm – a recursive procedure for maximization of the log likelihood function [25, 26]. However, there are serious obstacles on the way to developing an EM recursive algorithm for fitting a mixture model to the multi-component proteomic mass spectra. The first obstacle stems from the difficulty in setting initial conditions for EM iterations. Fitting a mixture model with a large number of components to data is difficult due to problems with setting appropriate initial conditions for the EM algorithm. The problem of setting the initial values of mixture parameters for the EM algorithm was researched numerous times in the literature [27-29]. However, the published approaches are practical only for mixtures with a relatively low number of components. When the number of components increases over 10-20, the precision of estimation of mixture parameters obtained with the use of the mentioned methods of initialization very rapidly decreases [30]. This makes the published methods inapplicable for mixtures with hundreds or even thousands of components encountered in spectra registered for complex proteomic mixtures (like serum or cancer tissue). An approach useful for setting initial components for the EM iterations dedicated to such spectra, was proposed in [18]. This approach applies an algorithm for detecting MS peaks as a first step and then sets initial mean values of components equal to detected locations of peaks. While the idea of using available information on locations of peaks of the spectrum is certainly reasonable and useful, the proposed approach still suffers from serious drawbacks: (i) EM iterations started with mean values of components positioned at MS peaks can still converge to undesired solutions due to imprecisions of initial values of component weights and standard deviations, (ii) the method is blind to "hidden" components, which are not identified (detected) by MS peaks, (iii) the method may require launching EM iterations at the full spectrum scale, which can be difficult for large datasets. The second obstacle is the size of the proteomic MS data. For very large datasets, with numbers of points along the m/z axis of a spectrum of the order of tens or even hundreds of thousands, executing (iterating) EM algorithm can be difficult due to large sizes of the necessary data structures and problems with the slow convergence.

In this paper we present a new algorithm for the Gaussian mixture modeling of protein mass spectra based on partitioning the MS signal into smaller fragments. The fragmented spectra are separately decomposed into mixture models. The obtained parameters of components for all fragments are then aggregated and used as the mixture model of the whole spectrum. The main idea of partitioning the MS signals into fragments by using "splitters", as well as other ideas of the elaborated algorithm, are described in detail in the "Methods" section of this paper. Partitioning the MS signal into fragments allows for overcoming both obstacles described in the previous paragraph. Both initializing and executing EM iterations is much easier for the smaller fragments of the MS signal than for the whole spectrum. Another advantage of partitioning the MS signals is the possibility of parallelizing the computations. Partition of the spectral signals into fragments is augmented by the use of an existing algorithm for peak detection for proteomic MS spectra.

We verify (prove) efficiency of the developed algorithm. In the first step of verification of our methodology we use our algorithm as a tool for improving peak detection in simulated mass spectra. We present comparisons of our algorithm of peak detection to the two peak detection algorithms of high efficiency published in the literature, MassSpecWavelet (based on continuous wavelet transform, CWT, approach) [3] and Cromwell (based on spectra differentiation) [2]. Comparisons are based on a large number of artificially generated datasets. We demonstrate the improvements achieved by using Gaussian mixture modeling. In the second step of verification of the methodology we are showing Gaussian mixture decompositions of real proteomic datasets [31, 32]. By visual inspections of Gaussian mixture models of the protein spectra and comparisons to locations of peaks detected by MassSpecWavelet and Cromwell we again demonstrate the ability of our algorithm to find "hidden" components of the MS signals.

# Methods

In this section we describe our algorithm for automatic, whole spectrum scale Gaussian mixture modeling (GMM) of proteomic mass spectra and for MS peak detection based on Gaussian mixture representation. We first introduce the notations for such spectra and their Gaussian mixture models. Afterwards, we first present the main idea of the algorithm and then the steps of the algorithm.

## The notation for MS spectral lines and their Gaussian mixture models (GMM). Scaling Gaussian mixture model of protein MS

A typical proteomic mass spectrum contains data on mass-to-charge (m/z) values of the registered ions, denoted by $x_n$ versus their abundances i.e., the numbers of counts from the ion detector, denoted by $y_n$, $n=1,2,...,N$. The number of data points in the spectrum is denoted by $N$. In real experiments the analyzed data-sets most often consist of more than one spectrum, multiple counts by $y_{mn}$, $m=1,2,...,M$ correspond to each point $x_n$ along the m/z axis, where $m$ denotes the index of the spectrum and $M$ is the number of the spectra. In some MS experiments it is possible for different spectra to be registered along different sets of points $x_n$.

As the model for proteomic mass spectra, we use the univariate Gaussian mixture probability density function of the form

$$f(x_n) = \sum_{k=1}^{K} \alpha_k f_k(x_n, \mu_k, \sigma_k) \quad (1)$$

where $K$ is the number of Gaussian components, $\alpha_k$, $k=1,2,...K$ are component weights (mixing proportions), which sum up to 1,

$$\sum_{k=1}^{K} \alpha_k = 1 \quad (2)$$

and $f_k(x_n, \mu_k, \sigma_k)$ denotes the probability density function of the Gaussian distribution.

$$f_k(x_n, \mu_k, \sigma_k) = \frac{1}{\sigma_k \sqrt{2\pi}} e^{-\frac{(x_n - \mu_k)^2}{2\sigma_k^2}} \quad (3)$$

In equations (1) and (3) $\mu_k$ and $\sigma_k$, $k=1,2,...K$, are means and standard deviations of the Gaussian components, respectively.

## Scaling

The mixture model (1) must be appropriately scaled. Due to finite sensitivity of the ion detector, numbers of counts in the average spectrum, $y_n$, correspond to ranges (intervals) in the m/z axis, $(x_n - \Delta_n/2, x_n + \Delta_n/2)$, where $\Delta_n$ is the width of the interval centered at $x_n$. In other words the data are binned and the numbers of counts $y_n$ are modeled by the multinomial probability distribution with probabilities given by areas of bins [26]. For real proteomic MS data bin widths $\Delta_n$ are changing with $n$; they are narrower for low $x_n$ and wider for high $x_n$. One can assume that bins are dense, i.e., their areas are well approximated by products of bin widths and values of the probability density functions at bin centers, which corresponds to the following model

$$y_n = \eta \Delta_n \sum_{k=1}^{K} \alpha_k f_k(x_n, \mu_k, \sigma_k) \quad (4)$$

The parameter $\eta$ in (4) is called the total ion current (TIC) [32]. From (4) the value of the total ion current $\eta$ is

$$\eta = \sum_{n=1}^{N} y_n \qquad (5)$$

We call model (4) the globally scaled model of the MS signal $y_n$ (due to its function of changing the scaling from probability densities to ion counts).

We are, however, more interested in using locally scaled models for the MS signal $y_n$. When we analyze only a small fragment of the spectral signal $n_{min} \leq n \leq n_{max}$, we can assume constant bin widths $\Delta_n = \Delta$. This allows for writing the locally scaled model in the following form

$$y_n = \sum_{k=1}^{K} w_k f_k(x_n, \mu_k, \sigma_k), n_{min} \leq n \leq n_{max} \qquad (6)$$

Where $w_k = s\alpha_k$, ($s = \eta\Delta$). The scale factor $s$ can be computed "locally".

$$s = \frac{\sum_{n=n_{min}}^{n_{max}} y_n}{\sum_{n=n_{min}}^{n_{max}} \sum_{k=1}^{K} \alpha_k f_k(x_n, \mu_k, \sigma_k)} \qquad (7)$$

## Main idea of the algorithm

As already stated in the Introduction section, the standard approach to estimating mixture parameters, $\alpha_k$, $\mu_k$ and $\sigma_k$, is by using expectation maximization (EM) recursive algorithm [25, 26]. Fitting the scaled mixture model (4) or (6) to spectral data $x_n$, $y_n$, $n=1,2,...,N$ is done by using an appropriate version of the EM algorithm, for binned data [26], described later in next subsection.

Fitting a mixture model to MS data by EM iterations at the whole spectrum scale is impractical (impossible) due to reasons described in the introduction. Therefore we have developed a method to decompose the MS signal into smaller fragments. Our method uses the property of the MS signal that after removing baseline (which is a wide component of the spectral signal) the remaining components are relatively narrow. The main idea of the algorithm is defining and modeling "splitters". A splitter is a fragment of an MS signal, which contains a "clear peak". An example of a clear peak and the related splitter in the protein MS signal is shown in Fig. 1. The position of the "clear peak" is marked by vertical red line in the upper plot and the splitter model is filled in red in the lower plot.

**Figure 1.** (A) Splitter segment – a fragment of the MS signal (black) around the clear peak detected in the MS signal (red vertical line close to m/z=6200). (B) GMM decomposition of the splitter-segment signal, splitter-segment signal (black), components of the Gaussian mixture model (green), mixture model signal (red), splitter computed on the basis of the clear peak in the upper plot (filled red).

In order to compute (estimate) the splitter signal model we "cut out" a fragment of the MS signal around the given (detected) clear peak (Fig. 1 A). This fragment of the MS signal is called a splitter-segment. We perform Gaussian mixture decomposition of the splitter-segment (Fig. 1 B). Cutting (truncating) the MS signal leads to possible errors in modeling. However, on the basis of our assumption of narrow components we expect that errors occur only close to boundaries of the splitting-segment and that they do not affect the model of the splitter (in the middle). Intuitively, in the vicinity of the clear peak the MS signal can be modeled by either only one or a small number of Gaussian components. These components are reliable parts of the decomposition of the MS signal into mixture of Gaussians.

For a given MS signal we need a set of splitters. Therefore, in the phase 1 of our algorithm, for a given MS signal, we search for a set of splitters by applying a heuristic procedure, which uses a peak detection algorithm as its first step. In principle any peak detection algorithm can be applied. In our implementation we used "mspeaks" function from Matlab Bioinformatics toolbox [33]. The heuristic procedure for searching for clear peaks (called also splitting peaks) is designed such that it returns a set of clear peaks, which are neither too close nor too far from each other and each of them is of sufficient quality (measured by the ratio of the peak height and heights of the neighboring lowest points of the MS signal). Then, as already stated, by using EM iterations we compute decompositions of splitting-segments (fragments of MS signals around each of the splitting peaks), as shown in Fig. 1, and we obtain models of all splitters signals.

Since models of splitters signals are reliable parts of Gaussian mixture decomposition of the MS signal, in the phase 2 of our algorithm we subtract splitters signals from the MS signal, which leads to splitting (partitioning) the whole spectrum into separate fragments, called segments. Then, segments are decomposed into Gaussian mixtures, again by using EM iterations. The idea of the phase 2 of our algorithm is illustrated in Fig. 2. In Fig. 2 A we present a fragment of the MS spectrum with two (neighboring) splitters. In Fig. 2 B we show the MS signal of the segment obtained by subtracting splitters models signals from the MS signal. In Fig. 2 C we show Gaussian mixture decomposition of the segment signal from the middle plot.

**Figure 2.** (A) Fragment of the MS signal (black line) with two neighboring splitters (filled red). (B) Segment (black line) resulting from subtracting the splitters signals from the MS signal. (C) Gaussian mixture decomposition of the segment signal, MS signal (black), components of the Gaussian mixture model (green), mixture model signal (red).

Finally, we aggregate all the computed GMM components into one set, which is a whole-spectrum mixture model of the MS signal.

# Steps of the algorithm

In this subsection we present a more detailed descriptions of successive steps of our GMM decomposition algorithm for MS signals.

## Baseline correction

We start the processing of the MS signal with the baseline correction (removal). In the proteomic MS data, the baseline can be quite precisely removed by using existing algorithms. In the implementation of our algorithms we are using the function "msbackadj" from the Matlab Bioinformatics Toolbox [33].

Baseline correction is an important step of the algorithm. After removing the baseline there is no need to introduce wide Gaussian components for its modeling. Therefore, the algorithm can be oriented towards searching for narrow components corresponding to protein/peptide species in the analyzed samples. By cutting-off negative values we ensure that the baseline corrected signal is non-negative (important for its mixture modeling). A baseline corrected MS signal is the data for the subsequent steps of the algorithm.

## Peak detection

We start the processing of the baseline-corrected spectrum by launching a peak detection procedure. There are many algorithms for peak detection of the MS signals in the literature. In our implementation we use "mspeaks" function from Matlab Bioinformatics toolbox [33], which provides estimates of both positions of spectral peaks and their widths (by the values of FWHH – full width at half high, returned by the "mspeaks" function). We divide each FWHH by the corresponding m/z value and we average over all detected peaks. The obtained value, proportional to the average coefficient of variation of Gaussian components, is used in the subsequent steps of the algorithm, for obtaining reliable estimates of sizes of splitting segments and segments, used in subsequent steps of the algorithm.

## Picking clear peaks and splitter-segments of the spectral signal

We go through all of the detected peaks and we compute quality of each peak, given by the ratio of the peak height and maximum of heights of the neighboring lowest points of the MS signal. Then we apply a heuristic procedure for picking clear peaks (splitting peaks). The requirements are that (i) each of the splitting peaks is of sufficient quality, (ii) distances between successive splitting peaks satisfy demands given by specified parameters (neither too close nor to far one from another). As already stated, we use average FWHH from the previous step to measure distances. For each clear peak (splitting peak) we define splitting-segment by cutting out a fragment of the MS signal around the splitting peak with suitably defined margins. For defining margins (sizes) of splitter-segments we again use average FWHH.

MS signals at the borders of the splitter-segments can sometimes assume quite high values, which can lead to substantial errors in mixture modeling. In order to reduce this effect we have designed a procedure of "warping down" of the splitting-segments. "Warping down" is a heuristic procedure of augmenting the analyzed splitter-segments by adding artificial parts of the signal outside their borders. The added parts assume values equal to signal values at the borders and vanish when the distance to the fragment increases.

## GMM decompositions of splitter-segments, computing splitters

Splitting-segments signals are decomposed into Gaussian mixture models by using EM iterations. Initialization and execution of EM iterations is described in the subsequent paragraph "Execution of EM iterations".

In the GMM model of the splitting-segment (Fig. 1 B) components close to borders may be unreliable, due to boundary effects. However, components close to the splitting peak are assumed free from disturbances coming from boundaries. Therefore, we pick up components, such that distances between their means and the position of the splitting peak are less than three standard deviations. These components result in the splitter signal model (filled in red in Fig. 1 B).

## Partitioning the spectrum into segments

We partition the baseline corrected MS spectrum into smaller fragments – segments, to later model each of them separately. Separated segment are obtained by subtracting splitter signals from the MS signal, as shown in Fig. 2 A,B.

## GMM decompositions of segments

Each of the segments is decomposed into a Gaussian mixture model by using EM iterations. The procedure for initialization and execution of EM iterations is described in the forthcoming paragraph "Execution of EM iterations".

## Execution of EM iterations

Parameters of the signal $y_n$, $n_{\min} \leq n \leq n_{\max}$, (6)-(7) corresponding either to a splitter-segment or to a segment are $w_k = s\alpha_k$, $\mu_k$ and $\sigma_k$, $k=1,2,...K$. EM recursions assume the following form (similar to standard EM iterations for mixtures) [18, 22, 26]

$$p(k|n) = \frac{\alpha_k f_k(x_n, \mu_k, \sigma_k)}{\sum_{\kappa=1}^{K} \alpha_\kappa f_\kappa(x_n, \mu_\kappa, \sigma_\kappa)} \quad (8)$$

$$\alpha_k = \frac{\sum_{n=n_{min}}^{n_{max}} p(k|n) y_n}{\sum_{n=n_{min}}^{n_{max}} y_n} \qquad (9)$$

$$\mu_k = \frac{\sum_{n=n_{min}}^{n_{max}} p(k|n) y_n x_n}{\sum_{n=n_{min}}^{n_{max}} p(k|n) y_n} \qquad (10)$$

$$\sigma_k^2 = \frac{\sum_{n=n_{min}}^{n_{max}} p(k|n) y_n (x_n - \mu_k)^2}{\sum_{n=n_{min}}^{n_{max}} p(k|n) y_n} \qquad (11)$$

In (8) $p(k|n)$ is the (estimated) conditional probability of measurements within the bin centered at $x_n$ belonging to (being generated by) $k$-th Gaussian component.

EM iterations (8)-(11) require specifying a method for setting initial values for parameters, $\alpha_k$, $\mu_k$ and $\sigma_k$. Appropriate initialization of EM is of critical importance for the convergence and quality of estimation. We are using here the algorithm for initialization of EM iterations, which applies dynamic programming partitions of the m/z values to estimate initial parameters. This algorithm shows certain advantages compared to other approaches.

Some additional assumptions (modifications) must be used for preventing a possible divergence of iterations. Mixture models fitted here to the MS signals are heteroscedastic (have unequal variances). For the case of unequal variances of components of the Gaussian mixture, the log-likelihood is unbounded [26, 34] which results in a possibility of encountering the divergence of EM iterations in computations. This problem is well known and there are several approaches published in the literature involving either a modification of the likelihood function [36] or introducing constraints on parameter values [35]. Here we prevent the divergence of EM iterations by simple constraint conditions concerning mixing proportions and component standard deviations.

We do not allow component standard deviations to fall below a certain threshold value. In other words we introducing the following additional operation in the iterations

$$\sigma_k = \max(\sigma_k, \sigma_{min}) \qquad (12)$$

The condition on mixing proportion involves removing too small components. If

$$\alpha_k < \alpha_{min} \qquad (13)$$

then the component $\alpha_k$, $\mu_k$, $\sigma_k$ is removed from the current mixture model, and further EM iterations resume with re-indexed components, re-scaled mixing proportions and decremented *K=K-1*.

Threshold values $\alpha_{min}$ and $\sigma_{min}$ are parameters of the algorithm. The default value $\alpha_{min}$ is assumed constant $\alpha_{min} = 10^{-3}$, while the value $\sigma_{min}$ depends on m/z; it is computed as 0.5*(m/z)*(average coefficient of variation of Gaussian components).

The last issue to resolve is estimating the number of components in the mixture model of the splitter-segment or segment. In order to estimate the number of components, *K*, EM iterations described above are launched multiple times with different *K*. Then the value *K* is chosen on the basis of some criterion function. A widely used method for estimating *K* is application of the Bayesian information criterion (BIC) [26, 37], which combines values of the log likelihood and a (scaled) penalty for the number of components. Here, for the choice of *K* we use an penalty index, $I_P$ with the structure analogous to BIC, namely

$$I_P = \Delta + K\check{n} \qquad (14)$$

In the above $\Delta$ is a scaled sum of absolute differences between the signal $y_n$ and its mixture model (the scaling factor is a value of TIC within the segment). We assume that $\varepsilon$ is a parameter to be chosen by the user. The default value is $\varepsilon = 0.002$ (obtained on the basis of computational experiments). The value of *K* is chosen on the basis of minimizing $I_P$ in (14) (over an assumed range of changes of *K*).

## Post-processing of the GMM model parameters

When GMM modeling is used for peak detection in MS signals, additional step of post-processing of the mixture model is necessary for improving its efficiency. Post processing includes two procedures, rejection of components corresponding to noisy elements of spectral signals and merging components.

Spectral signals decomposed by using GMM, apart from shapes corresponding to protein/peptide species (peaks) can contain noise and residuals of baseline signals. Existence of these disturbing parts of spectral signals result in obtaining components in the GMM model, which do not correspond to true peaks. We are filtering out using a cut-off value for weights. This cut-off value is obtained by using a maximum a posteriori rule to the two-component mixture model fitted to the set of all weights of components in the Gaussian mixture model.

Due to complexity of the mixture decomposition problem, it may happen that the obtained models contain components with similar values of means and variances (see examples of GMM decompositions in the Results section). For efficiency of the peak detection it might be reasonable to merge such similar components into one. The problem/need for merging similar Gaussian components was already encountered in practical applications of algorithms for decomposition of signals (datasets) into Gaussian mixtures. There are several papers devoted (or partly devoted) to methods of solving this problem e.g., [38].

Assume that there are two Gaussian components $\alpha_1 \mu_1 \sigma_1$ and $\alpha_2 \mu_2 \sigma_2$, which should be verified for being close enough to be replaced by one Gaussian component $\alpha, \mu, \sigma$. We compute differences $|\sigma_1 - \sigma_2|$ and $|\mu_1 - \mu_2|$. The merging threshold for standard deviations is assumed constant, equal to 0.05, and the merging threshold for means is denoted by

$$\text{MZ}_{\text{thr}} = |\mu_1 - \mu_2| \qquad (15)$$

We compute parameters $\alpha, \mu, \sigma$ assuming that they follow from maximum likelihood estimates based on observations generated by the mixture model $\alpha_1, \mu_1, \sigma_1, \alpha_2, \mu_2, \sigma_2$ which leads to estimates

$$\alpha = \alpha_1 + \alpha_2 \qquad (16)$$

$$\mu = \frac{\alpha_1}{\alpha} \mu_1 + \frac{\alpha_2}{\alpha} \mu_2 \qquad (17)$$

$$\sigma^2 = \frac{\alpha_1}{\alpha}\left(\mu_1^2 + \sigma_1^2\right) + \frac{\alpha_2}{\alpha}\left(\mu_2^2 + \sigma_2^2\right) - \mu^2 \qquad (18)$$

A Matlab implementation and exemplary data are available in supporting information file.

# Results

In this section we present some evaluations of the performance of our algorithm and comparisons to methods of analyses of MS signals based on spectral peaks. The presented results concern both simulated (low resolution) datasets, where true compositions samples corresponding to protein spectra are known, and real proteomic datasets of low and high resolution.

## Simulated data

First we apply our algorithm as a tool for peak detection for proteomic MS spectra. We compare our algorithm to two existing procedures for protein MS signals peak detection published in [2] and [3]. Our choice of the reference algorithms is based on the comparative studies [39-41] of algorithms for peak detection for the MS signals. The algorithm and associated computer program (R environment) published by Du and coauthors [3] was rated high in all comparisons studies as showing high sensitivity for peak detection with quite low false discovery rate. It is based on computing continuous wavelet transform (CWT) of the spectral signal, with the "Mexican Hat" mother wavelet function, and relating spectral peaks to the "ridge" lines in the parameter space. The algorithm developed by Coombes and coauthors in [2] (with publicly available implementation in the Matlab environment) was rated lower in comparisons [39-41]. When using this algorithm it is quite difficult to compromise between sensitivity of peak detection and false discovery rate. However, its advantage is that it uses natural ideas for peak detection, smoothing (with the use of wavelet functions) and differentiation of smoothed spectral signal. For the three compared algorithms we use the following abbreviations: MS-GMM – for our algorithm, CWT (continuous wavelet transform) – for the algorithm from [3] and CROM (Cromwell) – for the algorithm from [2].

Similarly to other studies devoted to comparisons of peak detection algorithms, [39-41], we use mass spectra, obtained with the use of the virtual mass spectrometer (VMS), [42], where the true positions of peaks in spectrum are known. We additionally change the structure of the simulated data by assuming different numbers of true peaks in the spectra, and we study their influence on the detection power of different algorithms. When applied to the simulated MS data, MS-GMM, CWT and CROM generate their lists of hypothetical spectral peaks. Hypothetical spectral peaks are compared to the true spectral peaks.

**Virtual mass spectrometer datasets**

Synthetic spectral datasets are obtained with the use of the VMS algorithm/tool [42] based on the physical principles underlying mass spectrometry instruments. This tool enables the generation of realistic virtual spectra with known underlying protein (peptide) compositions, and has already been widely used by many authors, [39, 41]. VMS signal contains the same parts as those (hypothetically) encountered in actual spectral signals, namely the true spectral signal consisting of a sum of overlapping Gaussian components (each corresponding to a protein or peptide species) multiplied by a random multiplicative factor adjusting for random amounts of proteins/peptides ionized and desorbed from each slide, a baseline signal and a zero mean Gaussian error with the variance given by a smooth function of m/z. For a given protein/peptide ion (i.e. spectral component) the we summarizes its distribution across samples by three quantities: its prevalence defined by the proportion of samples in the population containing the component, the mean and the standard deviation of corresponding peak intensity across samples that contain the component.

Our scenarios for simulating artificial datasets were similar to those previously applied by [39, 41]. By using the VMS algorithm we have generated five datasets with different true numbers of protein (peptide) species, 100, 150, 200, 250 and 300. Each dataset contained 100 spectral lines (samples) defined over the same equally spaced grid of 10000 m/z points over the range 2000-10000 Da, with a step 0.8 Da (which apparently corresponded to low resolution spectrum). In each dataset we have first randomly created a list of artificial components (proteins). For each component we randomly draw (i) its mass by using a uniform distribution supported over the range 2000-10000 Da, (ii) a value of the prevalence of this component from a beta distribution with parameters a=1, b=0.2, (iii) the component abundance from the right – shifted by 100 counts log-normal distribution with mean equal to 5 and variance equal to 1. For each sample we then determine whether the component is present in the sample by a random Bernoulli trial with the probability defined by the assumed value of the protein prevalence. If the Bernoulli trial returns 1 we generate the peak's intensity by drawing a random number distributed log-normally with mean $pe$ equal to the component's abundance and variance equal to $1.45\sqrt{pe}$. We then generate the position of the peak corresponding to the component (protein) using the normal distribution with mean $\mu$ equal to protein mass and standard deviation $\sigma = 0.001\mu$, which reflects the misalignment of the peak's position along the m/z axis between samples. The generated values of positions of peaks and their corresponding intensities are passed to the VMS algorithm, which generates an artificial spectrum containing the parts mentioned afore (the true spectral signal, baseline and noise). The default experiment parameters are used (mean initial velocity – 350 m/s, its standard deviation – 75 m/s, time between detector reads – 4e-9 s). Each synthetic spectrum includes a baseline and noise components. The baseline signal $y_B$ is modeled by using a formula with two exponential functions [39, 42]

$$y_B = b(1)e^{-\frac{x}{b(2)}} - b(3)e^{-\frac{x}{b(4)}} \tag{19}$$

where parameters *b(1)-b(4)* are randomly chosen from the list of estimated baselines for different cancers spectra [43]. The random noise component (signal) is modeled by using discrete ARMA (auto-regressive, moving average) model with 1 AR term and 6 MA terms [39].

Analogously to the studies [39-41] in each of the datasets (100, 150, 200, 250 and 300 true peaks) we base detection of peaks on the mean spectral signals. Mean spectra are computed by averaging the spectral signals over the same values of m/z coordinates [12]. Hypothetical spectral peaks generated by MS-GMM, CWT and CROM are compared to the true spectral peaks. If the hypothetical spectral peak lays within the ±0.3% range of the true position of the spectral peak the true peak is considered detected. Otherwise the true spectral peak is considered missed.

**Performance indexes**

We compute several performance indexes, useful to characterize/compare results obtained by different algorithms. The specificity index (defined by false discovery rate) is abbreviated by FDR. FDR is the number of peaks among those detected by the procedure which do not correspond to the true peaks, divided by the number of all peaks detected by the procedure. The sensitivity index is abbreviated by S. S is the number of true peaks detected by the procedure divided by the number of all true peaks in the sample. We also aggregate the performance measures FDR and S into one index, F1 (defined as the harmonic mean of 1-FDR and S)

$$F1 = \frac{2(1-FDR)S}{1-FDR+S} \qquad (20)$$

Obviously, higher values of F1 index imply better performance and lower values - poorer performance of the evaluated method. Finally, we also report the number of peaks detected by a peak detection algorithm.

All algorithms have free, tunable parameters, which should be chosen prior to their application. We use F1 index as base for tuning (optimizing) parameters of algorithms and we optimize performances of MS-GMM, CWT and CROM with respect to the average value of the F1 index over the ranges of their parameters analogously to [39-41].

In our algorithm MS-GMM we have adjusted one parameter, $MZ_{thr}$ (described earlier in the Methods section). Algorithms CWT and CROM include parameters, which can be tuned to data. For the CWT algorithm the adjustable parameters are Signal to Noise Ratio threshold (SNR), the scale range of the peak (peakScaleRange) and the minimal value of amplitude for the peak to be detected (ampTh). For the CROM algorithm the adjustable parameters are Signal to Noise Ratio threshold (SNR), a threshold for wavelet coefficients (threshold) and number of wavelets used in the transformation (L). Following recommendations of the authors and results from [39-41], for both algorithms we were tuning only two parameters, namely SNR and peakScaleRange for CWT, and SNR and threshold for CROM. Values of the third parameter in both algorithms were set constant ampTh=0, L=10. Optimization was performed separately for each dataset, with 100, 150, 200, 250, 300 true peaks, and involved averaged values of the F1 index. The obtained optimal values of parameters are reported in table 1 below.

**Table 1. Optimized values of parameters for algorithms MS-GMM, CWT and CROM**

| Algorithm | Parameters | Number of true peaks | | | | |
|---|---|---|---|---|---|---|
| | | 100 | 150 | 200 | 250 | 300 |
| **MS-GMM** | MZthr | 0.4 | 0.3 | 0.3 | 0.2 | 0.15 |
| **CWT** | SNR | 2 | 2 | 1 | 3 | 2 |
| | PeakScaleRange | 7 | 5 | 5 | 2 | 2 |
| **CROM** | SNR | 55 | 40 | 100 | 115 | 115 |
| | Threshold | 125 | 125 | 50 | 20 | 20 |

## Comparisons of performances of algorithms

The results of the comparison of performance indexes for the detection of peaks in the simulated datasets are presented in Fig. 3 where we show indexes of performances F1, FDR and S of the compared algorithms, as well as the numbers of (hypothetical) peaks detected by the algorithms.

**Figure 3. Performance indexes of the three peak detection algorithms applied for mean spectra in the simulated datasets**. (A) F1 score. (B) Sensitivity. (C) FDR. (D) No of detected peaks. Colors: MS-GMM – red, CWT – blue, CROM – green.

Performance indexes of all algorithms, MS-GMM, CWT and CROM show similar patterns of change. Consistently to results reported in the referenced comparison studies CROM achieves the lowest values of the F1 index for the whole range of true numbers of peaks in virtual spectra. Differences between algorithms are also seen when comparing numbers of the detected (hypothetical) peaks, shown in the Fig. 3 D. All algorithms underestimate the number of peaks.

Our algorithm MS-GMM exhibits the best performance in terms of values of the F1 index, for the whole range of values of true numbers of peaks in the virtual spectra. The possibility of tuning the parameter of this algorithm to achieve best compromise between sensitivity and FDR follows from highest values of sensitivities of our algorithm compared to other algorithms (Fig 3. B). Our algorithm is also closest to the truth when estimation the number of peaks in the spectral signal is considered.

Improvement of performance (high sensitivity) of peak detection achieved by using our Gaussian mixture model is obtained thanks to detection of "hidden" peaks in the spectral signals. This is illustrated in Fig. 4 below, where we have reproduced a small fragment (2900-3300 Da) of one spectral signal (with 200 true peaks). The plot includes positions of peaks detected by using CWT algorithm (blue asterisks) and components (peaks) detected by using our algorithm MS-GMM (green Gaussian curves). Along the x (m/z) line (bottom line of the plot) we have marked with circles all true peaks in the spectral signal (within the analyzed range) and we have additionally colored the circles depending on the detection status using the following code: detected only by MS-GMM method – red, detected only by CWT method – blue, detected by both MS-GMM and CWT – black and not detected by any of

algorithms – empty circle. One can see several examples of "hidden peaks", which have been detected thanks to the use of the Gaussian mixture model.

**Figure 4. Fragment of one virtual MS dataset (with 200 peaks, m/z range 2900-3300 Da). Comparison of MS-GMM and CWT.** MS signal (black), GMM model components (green), GMM model (red), peaks detected by CWT algorithm (blue asterisks). Positions of true peaks in the spectral signal are marked by circles symbols and detection status is depicted by colors: peak detected only by MS-GMM method (red), peak detected only by CWT method (blue), peak detected by both MS-GMM and CWT (black), peak not detected by any of algorithms (empty circle).

## Real proteomic datasets

We also show examples of GMM decompositions obtained with the use of our algorithm for real proteomic datasets. In the case of analyses of real datasets true compositions of samples are not known. For real proteomic spectral datasets comparisons of GMM modeling to methods based on spectral features defined by peaks can be done on the basis of indirect methods, e.g., on the basis of comparing efficiencies of spectral classifiers using different definitions of spectral features. We are, however, deferring such analyses to separate studies. Here, instead, we provide some technical comments on results of analyses of two real proteomic datasets concerning abilities of GMM modeling method to detect certain spectral components and concerning shapes of spectral signals encountered in real data.

### Low resolution dataset

The first dataset comes from published clinical study aimed at detection of colorectal cancer using serum peptidome profiling by MALDI-ToF mass spectrometry [31]. The dataset included 116 MALDI-ToF spectra of the low-molecular-weight fraction of serum proteome of cancer patients and healthy volunteers, each covering the m/z range 960–11,169 Da. Spectra were registered by Ultraflex MALDI-ToF spectrometer (Bruker Daltonics) working in the linear mode. The raw spectra contained approximately 45000 points along m/z axis. We used operations of averaging and binning described in the original paper, which resulted in "low resolution" spectra including approximately 10000 data points along m/z axis (~1 point per Da). Using our algorithm with the default settings resulted in computing the GMM model with 472 components. We did not apply any operations of post-processing of the GMM model. In Fig. 5 we show a short fragment of the mean spectrum of the data from [31] versus its GMM model. We also show, with blue asterisks, positions of peaks detected by using the CWT algorithm. From the plots in Fig. 5 one can again see several examples of parts of spectral signal, modeled by Gaussian components, not detected by peak detection algorithm (hidden peaks). In the right-hand part of the spectrum model one can see a low, wide component, which seems to result from some residue of baseline. Due to the fact that no post-processing was applied, there might be some excess of Gaussian components used to modeling. Therefore there are some of neighboring components, which can possibly be merged by the procedure described earlier.

**Figure 5. Fragment of the MS signal (within the range 1500-1650 Da) corresponding to the average spectrum of the serum peptidome for the data in [31]**. Spectral signal (black), GMM model signal (red), GMM components (green), peaks detected by using the CWT algorithm (blue asterisks).

### High resolution dataset

The second analyzed real dataset was generated in our team during characterization of head and neck cancer tissue proteome [32]. In this study a post-operative tissue sample was analyzed using MALDI Imaging Mass Spectrometry (MALDI-IMS). Tissue section processed with trypsin digestion was imaged with 50-100 μm raster using UltrafleXtreme MALDI-ToF spectrometer (Bruker Daltonics) working in the reflectron mode. Spectra were registered in the 800-4,000 Da range, which resulted in 20000 spectral signals, each containing 100000 data points along the m/z axis (i.e., ~30 points per Da, which could be considered as "high resolution" spectra). We have computed a mean spectrum (over 20000 signals) and we have decomposed it according to the GMM model, using our algorithm with the default settings, which resulted in obtaining 6216 components. In Fig. 6 we show a short fragment (1019-1030 Da) of mean spectrum versus the obtained GMM model. One can see a characteristic high resolution MS signal isotopic pattern with neighboring peaks occurring in the distance 1Da. One can also observe that isotopic parts of the spectral signal are asymmetric (right skewed). Typically, application of our algorithm results of modeling each of them by two Gaussian components, consistently to the theory in [19].

**Figure 6. Short fragment (1019-1030 Da) of the high-resolution mean spectrum corresponding to our own proteomic dataset of head and neck cancer tissues**. Spectral signal (black), GMM model signal (red), GMM components (green).

# Conclusion

We have managed to overcome the previously encountered difficulties and to develop an efficient method for the automatic whole spectrum decomposition of MS signals into Gaussian mixtures. The idea of the algorithm is based on partition of the spectral signal into separate fragments. The partition is obtained by defining "splitters" (fragments of an MS signal, which contain "clear peaks", as shown in Figs. 1 and 2). The possibility of the partition by splitters follows from properties of MS signals (with removed baselines). In the baseline corrected MS signals components are relatively narrow, which excludes long ranging overlaps. Separate segments are decomposed into GMM model by using EM iterations initialized with the use of the high efficiency algorithm [35]. Despite the multi – step design of our algorithm its ideas are simple. Partitioning of spectral signals allows for separate analyses leading to mixture models of sufficient precision. Aggregating results of decompositions of segments leads to Gaussian mixture model of the whole spectrum.

Separate decompositions of MS segments allow for estimation of whole spectrum Gaussian mixture models of MS signals of arbitrarily large sizes (proven by automatic analyses of high resolution spectra with numbers of m/z values of orders of hundreds of thousands). Separation also enables easy parallelizing of computations, which can be used to elaborate high efficiency computational environments based on multi-processor hardware systems.

When using the obtained mixture models for peak detection we have encountered a problem of selecting "peaks" from the set of mixture components. This problem arises because (i) some of the mixture components obtained in the iterative EM algorithm may not correspond to spectral peaks, (ii) application of the algorithm may result in obtaining two-component models of peaks. We have proposed a solution to this problem by a post-processing algorithm described in the Methods section, including a threshold value for component weights and a method for merging Gaussian components with a tunable parameter MZ-thr.

We have compared the peak detection method based on our GMM decomposition algorithm to two literature algorithms for peak detection (CWT, CROM), on the basis of artificially generated MS signals, and we have demonstrated its supremacy (Figs. 3 and 4).

Apart from improvement of the efficiency of peak detection demonstrated in this paper, there are also other areas of possible applications for an algorithm for the automatic, whole spectrum scale GMM decomposition of MS signal. Gaussian mixture modeling of MS signals can be (potentially) used as a tool for smoothing and de-noising spectral signals, for modeling and/or removing baselines in the spectra, for MS signals peak quantification, for MS signal (lossy) compression and for spectral deisotoping algorithms. Other applications can involve using mixture models for defining spectral features to be further used in construction of protein spectral classifiers.

# Supporting Information

**S1 File. A Matlab implementation of proposed algorithm and exemplary data.**

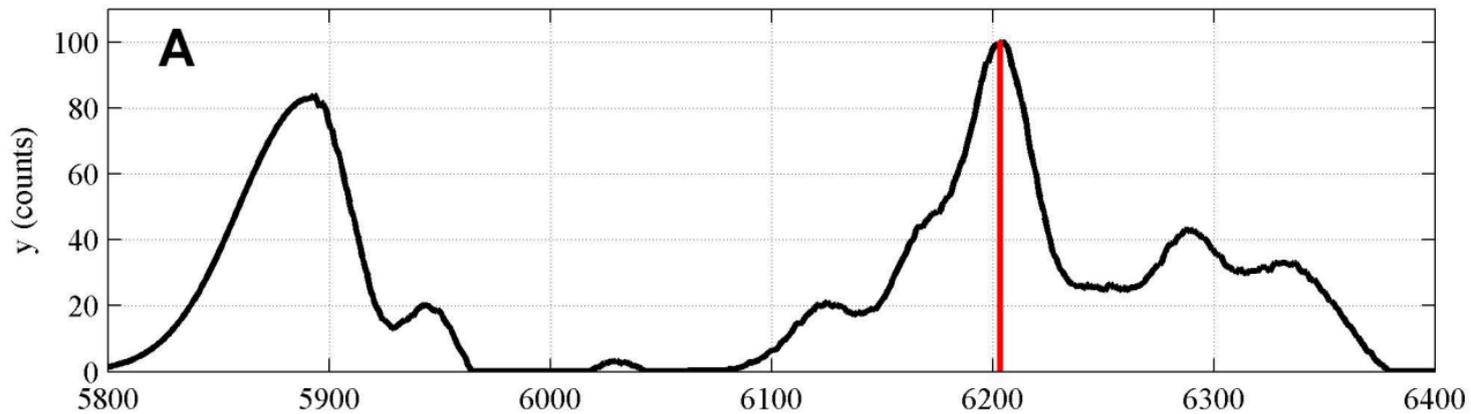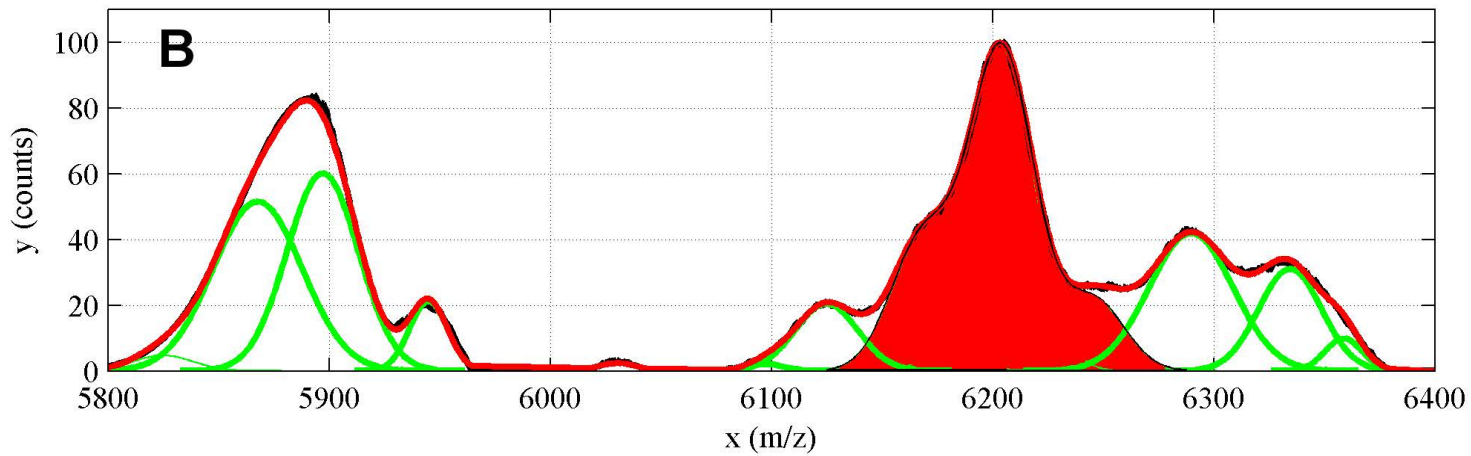

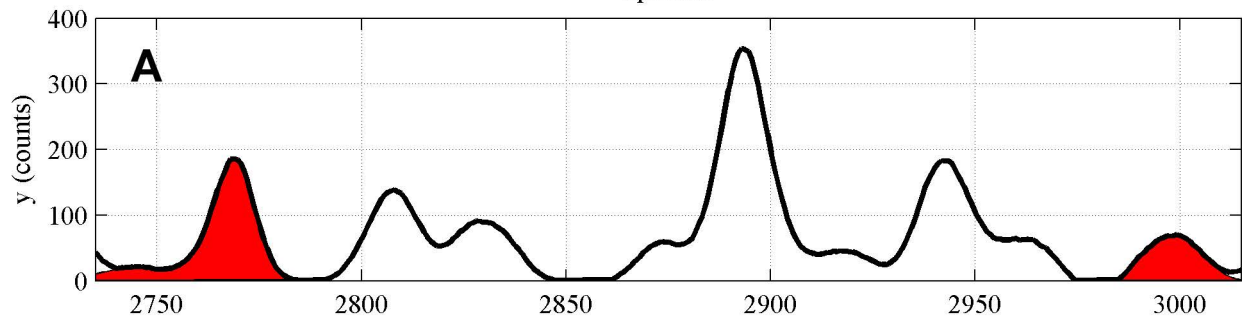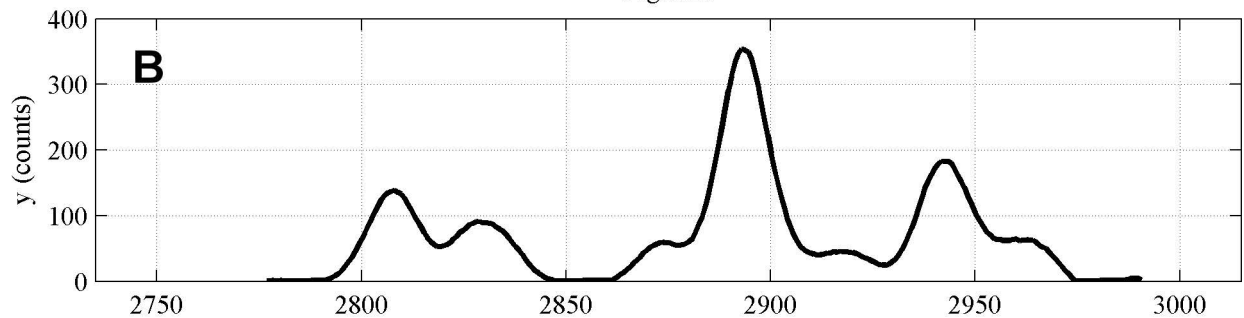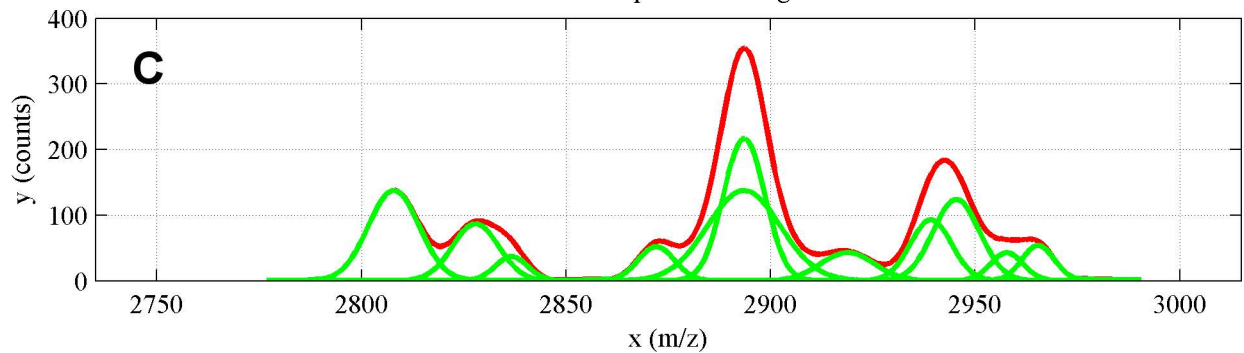

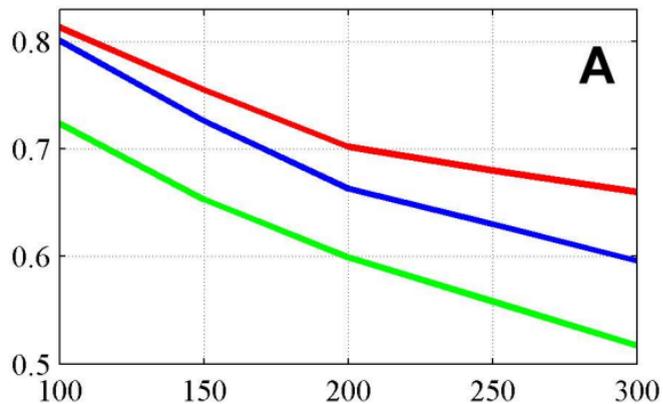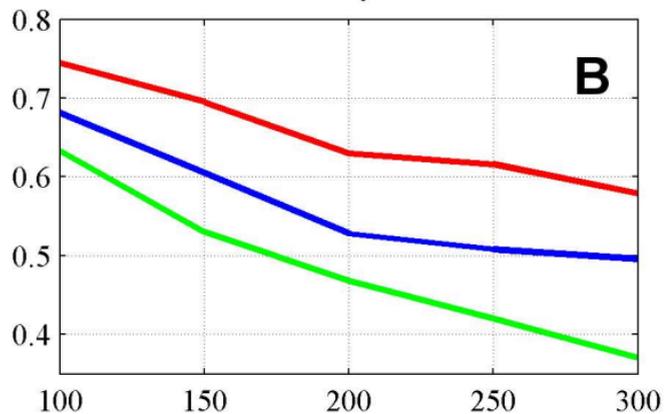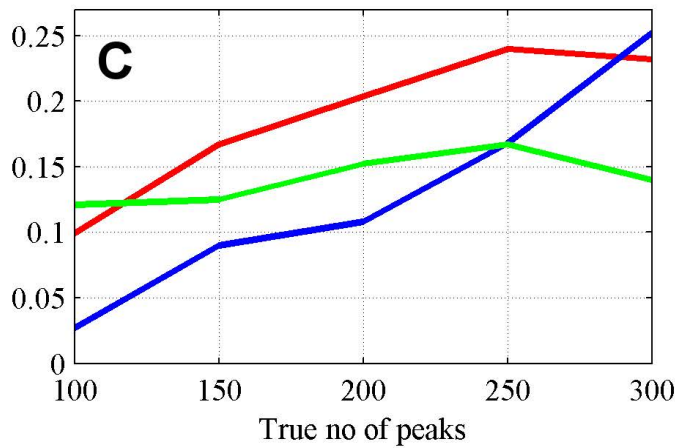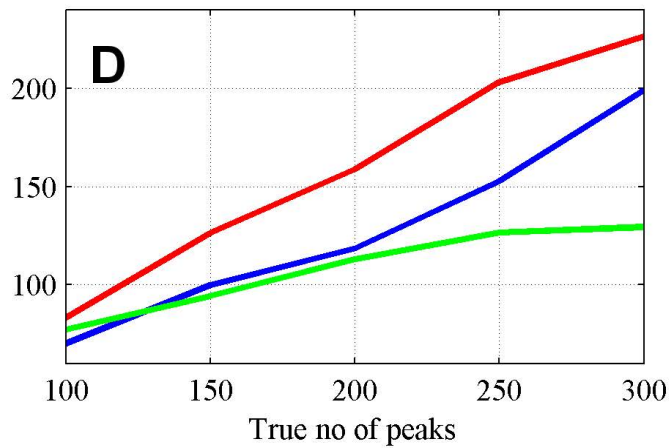

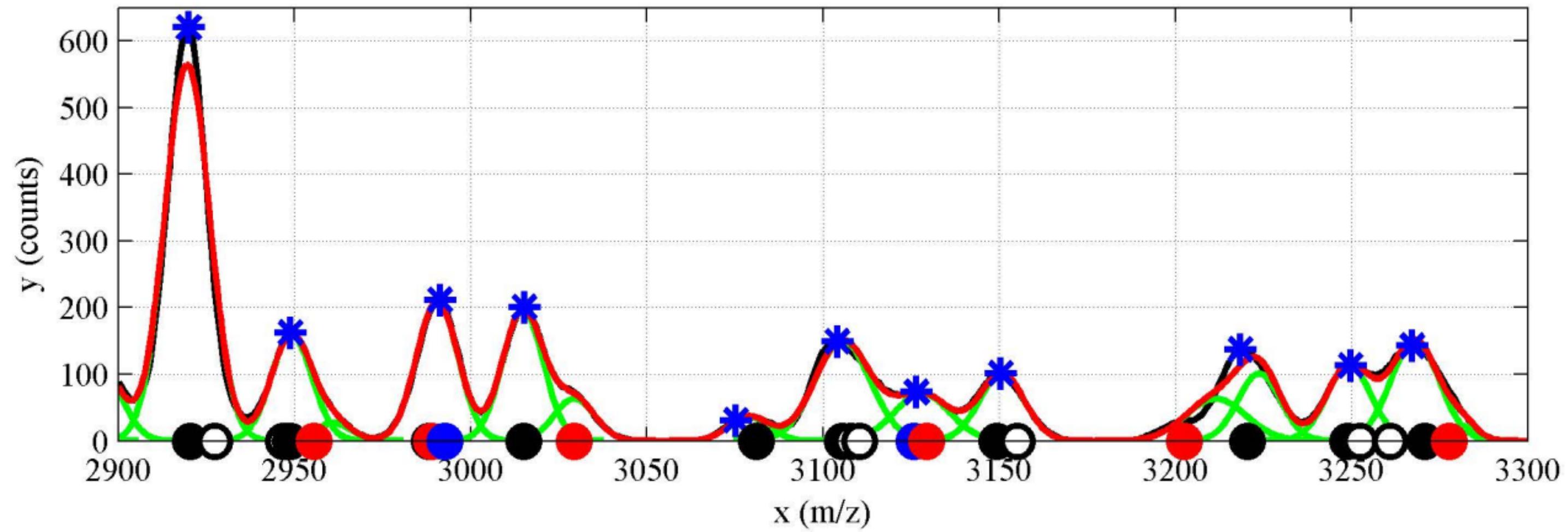

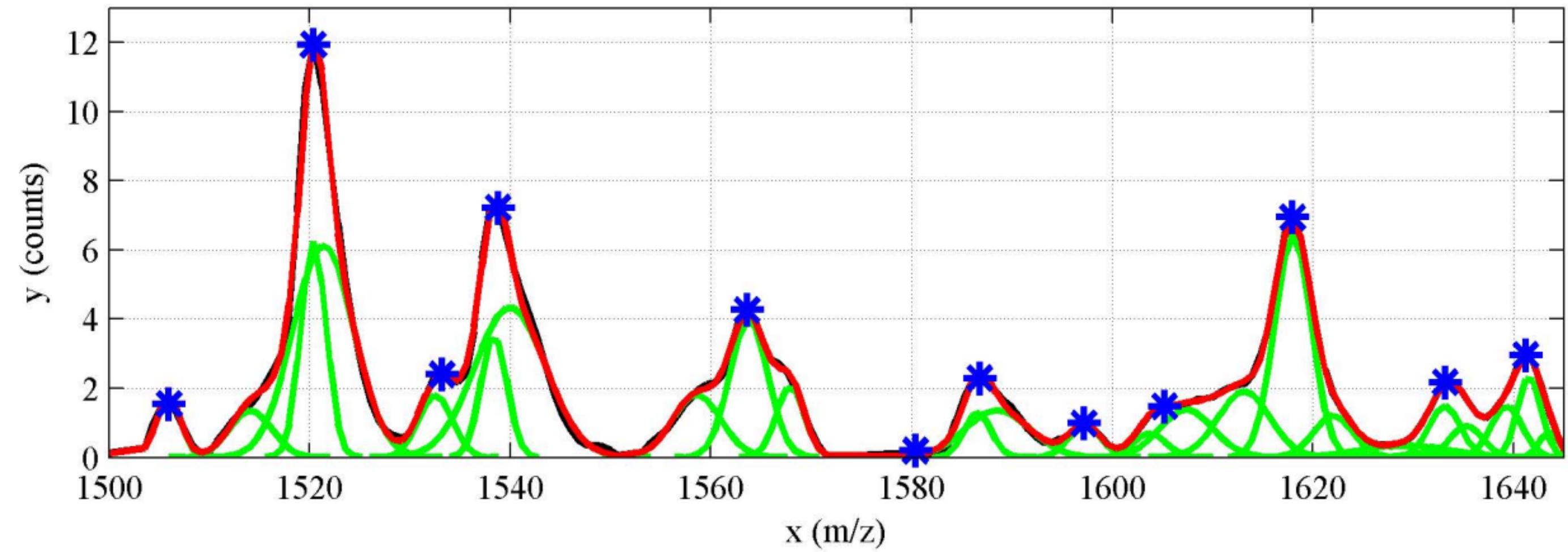

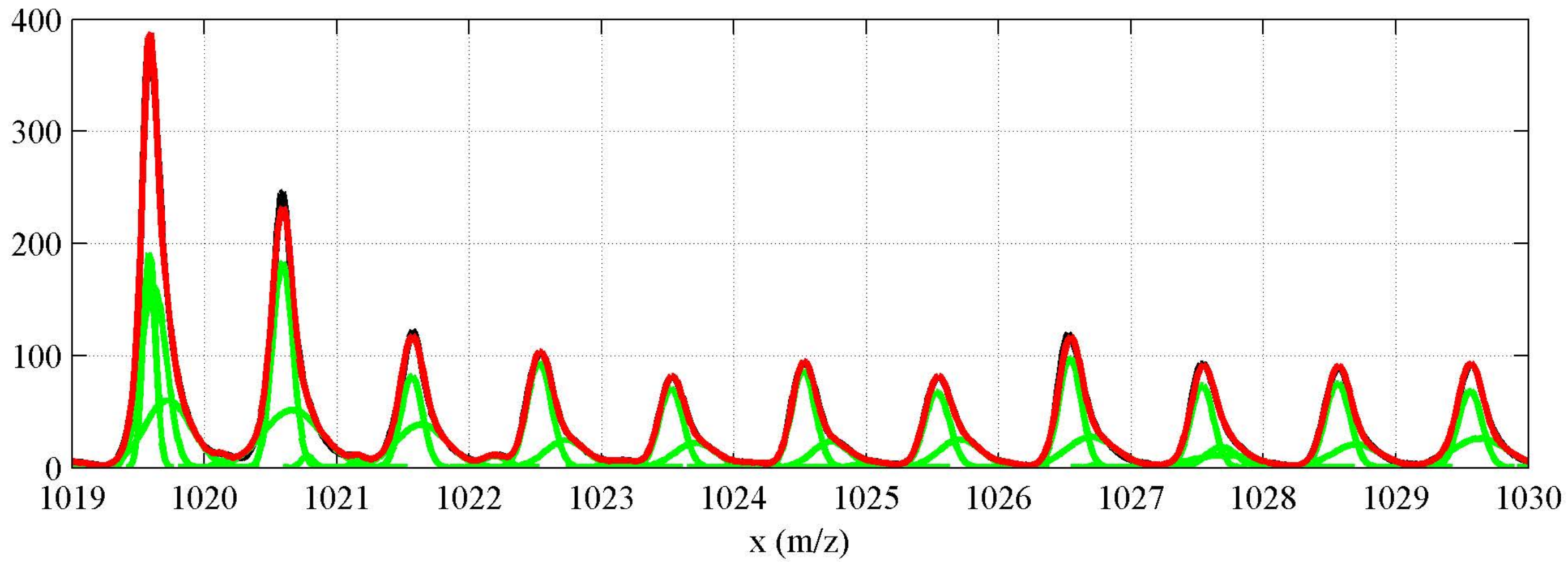